\documentclass[preprint,12pt]{elsarticle}
\usepackage{graphicx}% Include figure files
\usepackage{dcolumn}% Align table columns on decimal point
\usepackage{bm}% bold math
\usepackage{color}
\usepackage{caption}
\usepackage{graphicx, subfig}
\usepackage{epsfig}
\usepackage{hyperref}
\usepackage{ulem}
\hypersetup{
	colorlinks=true,
	linkcolor=blue,
	anchorcolor=blue,
	citecolor=blue}
\usepackage[dvipsnames]{xcolor}
\definecolor{green1}{rgb}{ 0.3882  ,  0.5961  ,  0.3882}
\usepackage[defaultcolor=Maroon]{changes}
\definechangesauthor[name={Tian-Xue Ma},color=BlueGreen]{MTX}
\newdimen\Lmargin
\newdimen\Rmargin
\Rmargin=\paperwidth \advance\Rmargin by-\Lmargin \advance\Rmargin by-\textwidth
\ifdim\Lmargin>\Rmargin \Rmargin=\Lmargin \fi
\def\Scentering{\leftskip=0pt plus 1fil minus \Rmargin
	\rightskip=\leftskip}

\usepackage{amssymb}
\usepackage{amsmath}
\journal{Materials Today Physics}

\begin{document}

\begin{frontmatter}



\title{Accelerated engineering of topological interface states in one-dimensional phononic crystals via deep learning}


%\author{} %% Author name
%
%%% Author affiliation
%\affiliation{organization={},%Department and Organization
\author[1]{Xue-Qian Zhang}%% Author name
\author[2]{Yi-Da Liu}
\author[3]{Xiao-Shuang Li}
\author[1]{Tian-Xue Ma \corref{cor1}}%% Author name
\ead{matx@bjtu.edu.cn}
\author[1,2]{Yue-Sheng Wang \corref{cor1}}%% Author name
\ead{yswang@bjtu.edu.cn}
\author[4]{Zhuo Zhuang}%% Author name

\address[1]{Department of Mechanics, School of Physical Science and Engineering, Beijing Jiaotong University, Beijing, 100044, China}
\address[2]{Department of Mechanics, School of Mechanical Engineering, Tianjin University, Tianjin, 300350, China}
\address[3]{College of Civil Engineering and Architecture, Hebei University, Baoding, 071002, China} 
\address[4]{School of Aerospace Engineering, Tsinghua University, Beijing, 100084, China}
\cortext[cor1]{corresponding author}

%% Abstract
\begin{abstract}
Topological interface states (TISs) in phononic crystals (PnCs) are robust acoustic modes against external perturbations, which are of great significance in scientific and engineering communities. However, designing a pair of PnCs with specified band gaps (BGs) and TIS frequency remains a challenging problem. In this work, deep learning (DL) approaches are used for the engineering of one-dimensional (1D) PnCs with high design freedoms. The considered 1D PnCs are composed of periodic solid scatterers embedded in the air background, whose unit cell is divided into a matrix with $32 \times 32$ pixels. First, the variational autoencoder is applied to reduce the dimensionality of unit cell images, allowing accurate reconstruction of PnC images with different numbers of scatterers. Subsequently, the multilayer perceptron and the tandem neural network are used to realize the property prediction and customized design of 1D PnCs, respectively. The correlation coefficients for the property prediction and inverse design are more than 97$\%$. The unit cell images of 1D PnCs with specific BG properties could be successfully and instantaneously designed. Importantly, the implementation of a "one-to-many" design of PnC pairs with specific TIS frequencies is realized. Furthermore, the reliability and robustness of the constructed networks are confirmed by randomly specifying the design targets as well as the experimental verification. This study demonstrates the broad application prospects of DL approaches in the field of PnC design and provides new ideas and methods for the intelligent design of artificially functional materials.
\end{abstract}


%\item The topological images of phononic crystal unit cells with highly design freedoms;
%\item Co-design of band gap boundary frequencies and topological properties;
%\item Design directly for topological interface state frequencies;
%\item Implement a "one-to-many" design for the same topological interface state frequency;
%\item Network prediction $10^{5}$ faster than finite elements, millisecond-level design.

%% Keywords

\end{frontmatter}



\section{Introduction}
Phononic crystals (PnCs) are a new type of periodically functional materials that have a unique ability to manipulate the propagation of acoustic or elastic waves. In general, a band structure is used to evaluate the wave characteristics of PnCs, as it determines the allowed and forbidden frequency ranges for wave propagation, the so-called passbands and band gaps (BGs). The regulation of BGs enables various potential applications, such as noise barriers \cite{radosz2019acoustic}, wave filters \cite{muhammad2021phononic}, waveguides \cite{khelif2003two, khelif2004guiding} and negative refraction \cite{jia2023optimization}. Stimulated by topological insulators in condensed matter physics, there is growing interest in studying topological phases of PnCs. PnCs with topologically non-trivial characteristics could exhibit topologically protected states. For example, topological interface states (TISs) emerge at the interface between two crystals with different topological properties \cite{he2021inverse}. Importantly, the TISs in PnCs are robust against defects and can propagate through sharp corners without back-scattering, enabling low-loss and high-efficiency sound wave transmission \cite{singh2021spin, luo2021moving}. This characteristic is highly significant for the design of high-performance acoustic devices. By breaking the time-reversal symmetry \cite{wang2015topological, chen2016tunable}, a phenomenon analogous to the quantum Hall effect can occur. Additionally, the band inversion through breaking lattice symmetries could lead to the generation of acoustic topological insulators comparable to the quantum valley Hall effect \cite{yan2018chip, xue2024three}. Furthermore, in two-dimensional (2D) systems, acoustic topological insulators could also be achieved through the quantum spin Hall effect \cite{he2016acoustic, chen2024elastic}. The above-mentioned approaches can lead to the opening of Dirac cones and the formation of a BG with topologically non-trivial characteristics. In addition, for one-dimensional (1D) PnCs, the topological properties can be characterized by the Zak phase \cite{xiao2015geometric}. Taking the Su-Schrieffer-Heeger (SSH) model as an example, the topological phase transition is achieved by changing the inter- and intra-cell hopping strengths \cite{yin2018band, zhou2019topological, tang2023topological}. The TISs in 1D systems are localized states with high robustness, which can be utilized for functions such as energy focusing \cite{MA2022101578}, filtering \cite{ortiz2021topological}, and sensing. However, customizing PnCs with specific BGs and desired topological properties is an intractable problem. Traditionally used trial-and-error or empirical methods are time-consuming and computationally intensive. 

Deep learning (DL) is one kind of neural networks that can recognize complex data and make accurate predictions. It has been widely used in computer vision \cite{jiang2023review, ikromovich2023facial}, speech recognition \cite{kheddar2024automatic}, and natural language processing \cite{khan2023exploring}. Likewise, DL is promising for challenges of classical waves and thus can be employed to predict and design dispersion relations \cite{jiang2022dispersion}, BGs \cite{li2020designing, liu2022inverse}, and sound absorption spectra \cite{zhang2023learning} of various PnCs. For the inverse design of topological insulators, He et al. \cite{he2022machine} and Long et al. \cite{long2019inverse} used DL tools to design 1D PnC beams and 1D layered photonic crystals, respectively. The corresponding crystals with different Zak phases were successfully designed, supporting the generation of TISs at the structural interfaces. However, their design domain is restricted to several parameters, lacking the design of geometrical topology of a structural image. That is, the freedoms of design for a PnC image are significantly limited. Notably, topology optimization can also be used for the inverse design of PnCs with particular wave properties. Wang et al. \cite{WANG2024129571} achieved the design of TIS frequencies by optimizing the geometrical topology of 1D PnCs. However, their work realizes the TISs by designing the transmission spectra, and there is a lack of direct prediction and design of TIS frequencies. Moreover, it is pointed out that topology optimization requires re-optimization every time once the design objective is changed. In contrast, after the training of DL model is completed, it can simultaneously handle task with multiple targets. Also, it is possible to combine the above-mentioned two approaches for the inverse design of PnCs with expected wave properties \cite{mao2023multi, KUDELA2023110636}. However, there is still a lack of research on precise engineering of TISs in complex PnCs using DL models.  

In this paper, we explore the potential of DL tools for accelerated engineering of 1D topological PnCs with high design freedoms. The paper is organized as follows. The PnC geometry, the DL models, and the calculation methods used in this paper are described in Section \ref{section2}. Section \ref{section3} presents the results and discussion, including the dimensionality reduction of PnC geometry, the prediction of wave characteristics, the inverse design of PnCs, as well as the experimental validation. Finally, the conclusion is drawn in Section \ref{section4}.

\section{Model and methods} \label{section2}
\subsection{PnC geometry}

\begin{figure}[htbp]
	\centering
	\includegraphics[scale=1]{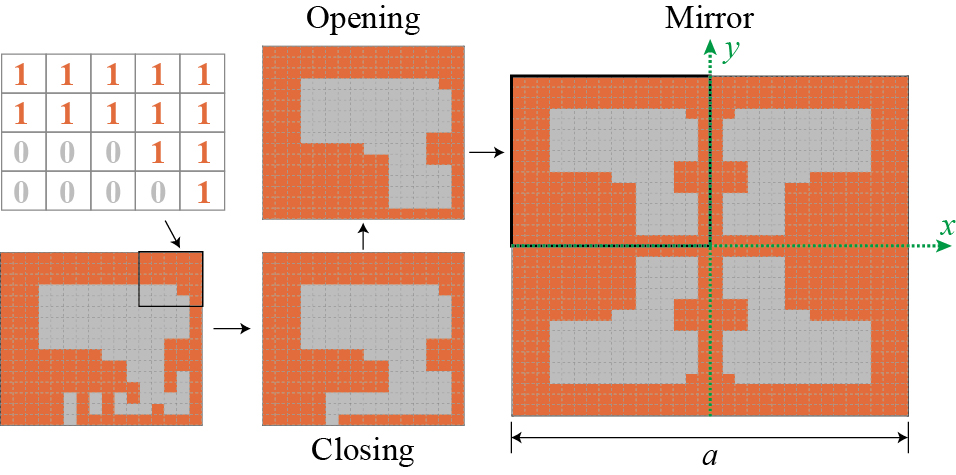}% Here is how to import EPS art
	\caption{\label{Fig.1}Schematic diagram for establishing the PnC unit cell, where the randomly generated image with $16 \times 16$ pixels is mirrored in the $x$- and $y$-axes after the closing and opening operations.}
\end{figure}

To construct the 1D PnCs, solid scatterers (in gray color) are arranged periodically along the $x$-axis in a waveguide filled with air (in orange color), as schematically shown in Fig. \ref{Fig.1}. Throughout this work, it is assumed that the unit cell of PnCs possesses the $C_{2\textrm{v}}$ symmetry. For the purpose of encoding, the PnC unit cell is divided into a $32 \times 32$ binary matrix, where 0 and 1 represent the scatterers and the air domain, respectively. According to the $C_{2\textrm{v}}$ symmetry, we first create a binary image of 16 by 16 pixels that represents the 1/4 PnC unit cell. Then, small holes are filled using the closing operation, and narrow parts are removed by the opening operation, ensuring that each scatterer area is smooth and four-connected. Finally, the 1/4 unit cell is mirrored in the $x$- and $y$-axes, respectively. Notably, we also impose the constraints on the air domain to ensure that it is simply connected and directly connected to the neighboring unit cells (along the $x$-axis). The generation process for PnC unit cells is plotted in Fig. \ref{Fig.1}. 

\subsection{Band structure and Zak phase}

The governing equation for the propagation of linear acoustic waves is
\begin{equation}
	%\label{eq1}
	\begin{split}
		\ \bigtriangledown \cdot [\frac{1}{\rho(\textbf{r})} \bigtriangledown p(\textbf{r},\textbf{k})] = - \frac{1}{\rho (\textbf{r}) c^{2}(\textbf{r})} \omega^{2} p(\textbf{r},\textbf{k}),
	\end{split}
\end{equation}
where $p(\textbf{r},\textbf{k})$ represents the acoustic pressure field; $\rho(\textbf{r})$ represents the mass density; $c(\textbf{r})$ is the sound velocity; $\omega$ is  the angular frequency; $\textbf{r}$ and $\textbf{k}$ denote the position vector and the wave vector, respectively. According to Floquet-Bloch theorem, the pressure field in any PnC can be formulated as
\begin{equation}
	%\label{eq1}
	\begin{split}
		\ p(\textbf{r}) = p_{\textbf{k}}(\textbf{r})e^{i(\textbf{k} \cdot \textbf{r})} ,
	\end{split}
\end{equation}
where $p_{\textbf{k}}(\textbf{r})$ is a periodic function. Besides, the PnC unit cell satisfies the following Bloch's periodic boundary:
\begin{equation}
	%\label{eq1}
	\begin{split}
		\ p(\textbf{r}+\textbf{a}) = p(\textbf{r})e^{i(\textbf{k} \cdot \textbf{a})},
	\end{split}
\end{equation}
where $\textbf{a}$ is the lattice vector. In this paper, the numerical calculations are performed by the finite element method (FEM) using COMSOL Multiphysics. By solving the governing equation using the FEM, one can obtain
\begin{equation}
	%\label{eq1}
	\begin{split}
		\ (\textbf{K}-\omega^{2}\textbf{M})\textbf{P}=0,
	\end{split}
\end{equation}
where $\textbf{K}$ is the stiffness matrix; $\textbf{M}$ is the mass matrix; $\textbf{P}$ represents the eigenvector of pressure field. The dispersion relations of a 1D PnC can be calculated by sweeping the wave vector $\textbf{k}$ along the $\Gamma$-X direction of the first irreducible Brillouin zone (IBZ) of the 1D lattice, i.e., from $0 \sim \pi/a$.

\begin{figure}[htbp]
	\centering
	\includegraphics[scale=1]{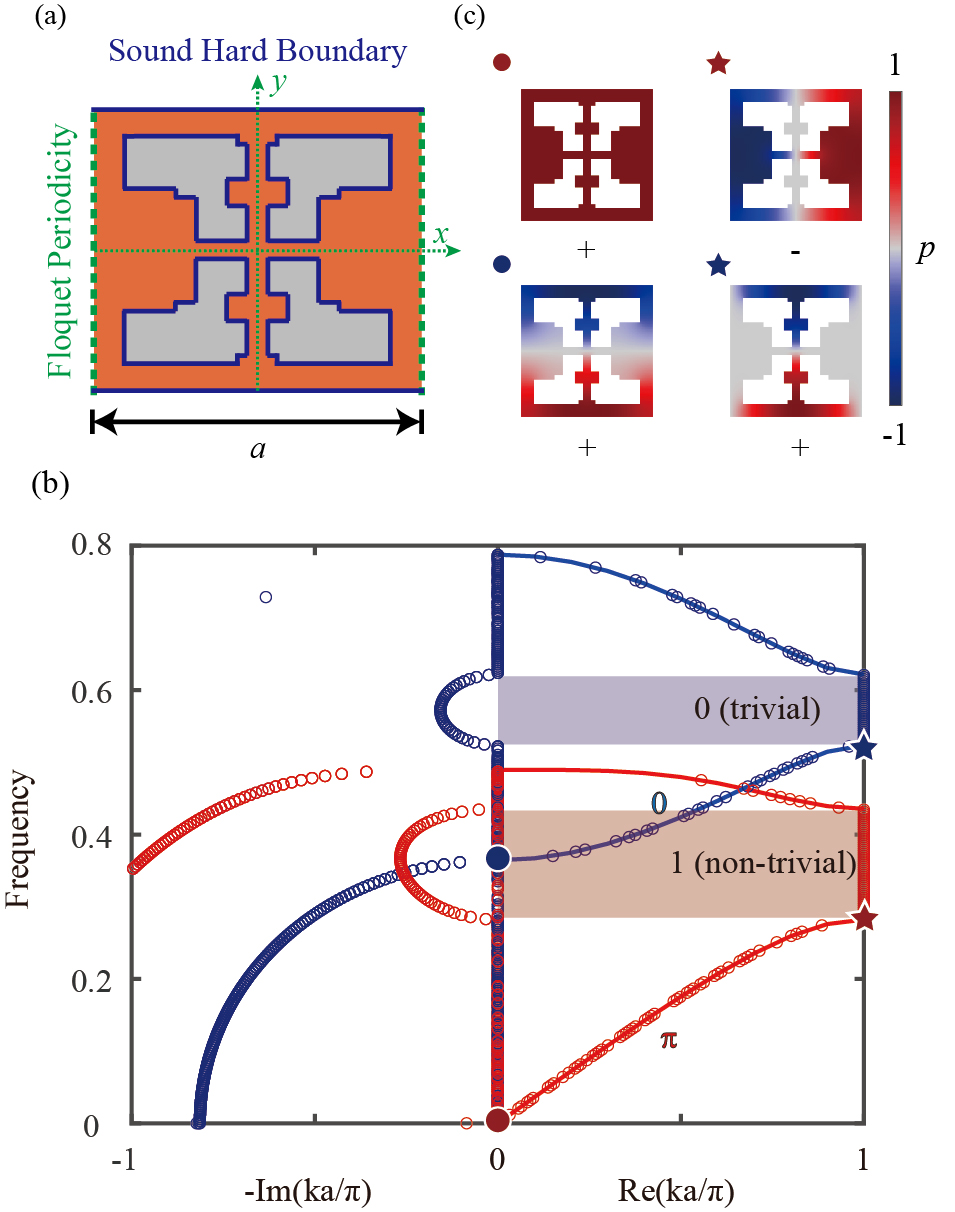}% Here is how to import EPS art
	\caption{\label{Fig.2}(a) Schematic illustration of a PnC unit cell in the dataset, where the air background and the solid scatterers are displayed by the orange and gray colors, respectively. The sound hard and Floquet periodic boundaries are denoted by the blue solid lines and green dashed lines, respectively. (b) Band structure of the 1D PnC in (a), where the solid lines and the hollow dots represent the real and complex dispersion curves, respectively. The red and blue colors correspond to the symmetric and anti-symmetric acoustic modes with respect to the $x$-axis, respectively. (c) Eigenmodes at the highly symmetric points $\Gamma$ and X marked in (b).}
\end{figure}

Here, one PnC unit cell in the dataset is illustrated in Fig. \ref{Fig.2}(a). Notably, due to its large acoustic impedance the solid material is rigid for the airborne sounds. In the numerical calculations, the sound velocity and the mass density of air are $c$ = 343 m/s and $\rho$ = 1.21 kg/m$^{3}$, respectively. The Bloch's periodic boundary conditions are applied in the $x$-direction [indicated by the green dashed lines in Fig. \ref{Fig.2}(a)], while the other boundaries are set as the sound hard boundaries [indicated by the blue solid lines in Fig. \ref{Fig.2}(a)]. The real band structure of the 1D PnC is calculated [see the solid lines in Fig. \ref{Fig.2}(b)], in which the frequency $f$ is normalized by $c/a$ with $a$ being the lattice constant. Interestingly, as the considered 1D PnCs are symmetric with respect to the $y$-direction, they support both the symmetric and anti-symmetric modes, as plotted in Fig. \ref{Fig.2}(c). In the band diagram, the red and blue colors represent the symmetric and anti-symmetric modes, respectively. It is important to note that these two modes can hardly couple, especially in low frequencies. Concerning this point we also calculate the complex band structure using the $\omega$-\textbf{k} approach \cite{laude2009evanescent}, as indicated by the hollow circles in Fig. \ref{Fig.2}(b). From the complex band structure it is observed that, for the corresponding first two bands, the symmetric mode does not interact with the anti-symmetric mode. Therefore, in this study, we focus on the characteristics of the first-order BGs (between the first and second bands) for the symmetric and anti-symmetric modes, respectively.

For a 1D PnC with the mirror symmetry, the topological property of each isolated band can be expressed by the Zak phase $\theta$ \cite{zhang2019subwavelength, ma2022topological}:
\begin{equation}
	%\label{eq1}
	\begin{split}
		\frac{\theta}{\pi} = \frac{1}{2} [\eta(k_{x}=0) - \eta(k_{x}=\frac{\pi}{a})] : \mathrm{modulo 2},
	\end{split}
\end{equation}
where $\eta$ represents the mode parity at the highly symmetric points. The Zak phase is either 0 or $\pi$, signifying the topologically trivial and non-trivial phases of the considered band, respectively. As shown in Fig. \ref{Fig.2}(c), the positive and negative signs indicate the parity of eigenmodes with respect to the $y$-axis. Thus, the Zak phases of the first-order symmetric and anti-symmetric bands are $\pi$ and 0 respectively. The topological property of a BG can be evaluated by the following relation:
\begin{equation}
	%\label{eq1}
	\begin{split}
		\mathrm{sgn}[\zeta^{(n)}] = (-1)^{n+1} \mathrm{exp}(i\sum_{m=1}^{n}\theta_{m}),
	\end{split}
\end{equation}
where $n$ represents the BG order. Thus, the first-order symmetric and anti-symmetric BGs are characterized the corresponding first bands, in which the negative sign $(\zeta<0)$ and the positive sign $(\zeta>0)$ imply topologically non-trivial and trivial, respectively. Notably, in the dataset, these properties are labeled by 1 (non-trivial) and 0 (trivial).

\subsection{Deep learning models}

\begin{figure*}[htbp]
	\Scentering
	\includegraphics[scale=1]{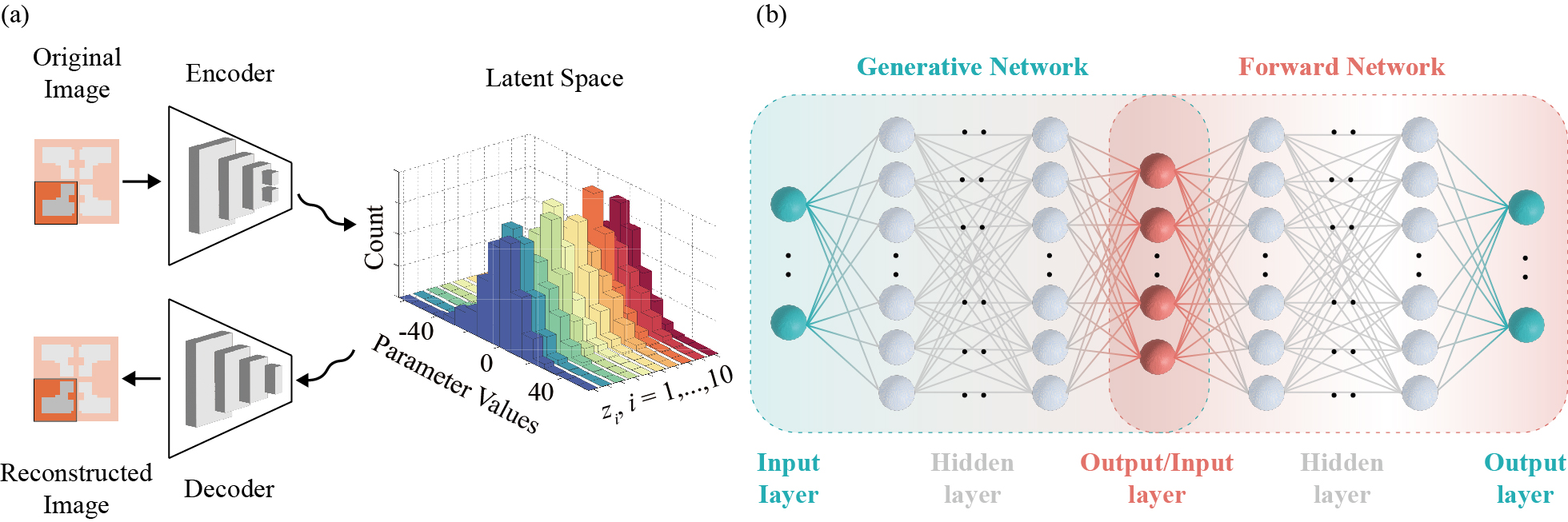}% Here is how to import EPS art
	\caption{\label{Fig.VAE_Z_MLP_TNN}Schematic diagrams of the VAE (a) and the TNN (b). The TNN is comprised of a generative network and a trained forward network (i.e., the MLP).}
\end{figure*}

The VAE is a generative artificial intelligence (AI) algorithm that consists of two main components: an encoder and a decoder. The encoder extracts the features from high-dimensional data and then maps them to a low-dimensional space, while the decoder reconstructs the low-dimensional data back to the input space (i.e., the high-dimensional data). Since the PnC unit cells possess the $C_{\textrm{2v}}$ symmetry, the VAE is employed to map one-quarter of unit cells ($16\times16$ pixels) to a low-dimensional vector of $1 \times 10$ dimensions, as shown in Fig. \ref{Fig.VAE_Z_MLP_TNN}(a). The loss function comprises the Kullback-Leibler (KL) divergence loss and the reconstruction loss. The former measures the difference between the latent space and the standard normal distribution, while the latter assesses the difference between the original and reconstructed samples. The loss function is defined as follows:
\begin{equation}
	%\label{eq1}
	\begin{split}
		\ L_{\mathrm{VAE}} = -\frac{1}{M} \sum\limits_{m=1}^{M} \{
		\frac{1}{2} \sum\limits_{i=1}^{10} [ 1-(\alpha_{i}^{(m)})^{2} - (\sigma_{i}^{(m)})^{2} + 2\mathrm{ln}(\sigma_{i}^{(m)})] \}
		\\
		-\sum_{n=1}^{N} [ x_{n}^{(m)} \cdot \mathrm{ln}(\widehat{x}_{n}^{(m)})+(1-x_{n}^{(m)})\cdot \mathrm{ln}(1-\widehat{x}_{n}^{(m)})
		],
	\end{split}
\end{equation}
where $M$ is the minibatch; $\alpha$ and $\sigma$ represent the mean and standard deviations of the latent vector, respectively; $N$ is the dimension of design variables for the original PnC unit cells, with the size of $16\times16$; $x$ and $\widehat{x}$ represent the pixel values of the original and reconstructed unit cells, respectively. The latent vector $\textbf{z}$ is generated by combing a random noise $\varepsilon$ with the mean and standard deviations:
\begin{equation}
	%\label{eq1}
	\begin{split}
		\ \mathrm{z}_{i} = \alpha _{i} + \varepsilon \sigma _{i}, i=1,2,...,10 ,
	\end{split}
\end{equation}
where $\varepsilon$ is sampled from a standard normal distribution. To train the VAE effectively, we use Adam as the optimizer, with a learning rate (LR) of 0.0001 and a batch size of 1024. 

The MLP is an artificial neural network composed of an input layer, a number of hidden layers, as well as an output layer. Each layer contains multiple neurons that perform a weighted summation of signals from the previous layer and apply an activation function to produce nonlinear transformations. This enables the MLP to approximate any nonlinear function. Herein, the MLP is used to predict the physical properties of the 1D PnCs, as illustrated by the forward network on the right part of the TNN in Fig. \ref{Fig.VAE_Z_MLP_TNN}(b). In details, the properties of the first-order BGs for symmetric and anti-symmetric modes (i.e., boundary frequencies and topological properties), as well as the frequencies of TISs are predicted via the MLP framework. The details of calculating the frequencies of TISs within finite-sized PnC structures are given in \ref{A}. Specifically, the three MLPs for predicting the BG frequencies, BG topological properties and TIS frequencies are named as follows: MLP$_{\mathrm{boundary}}$, MLP$_{\mathrm{topo}}$, and MLP$_{\mathrm{TIS}}$. Besides, it is worth noting that the latent vectors and predicted targets are used as the inputs and outputs in the forward networks (i.e., the MLPs), respectively. The three loss functions are as follows:
\begin{equation}
	\begin{array}{l}
		L_{\mathrm{boundary}} = \frac{1}{M} \sum_{i=1}^{M} \sum_{j}^{4} |f_{ij} - \widehat{f}_{ij}|,
		\\[1em]
		L_{\mathrm{topo}} = -\frac{1}{M} \sum_{i=1}^{M} [t_{i}\mathrm{log}(\widehat{t}_{i}) + (1-t_{i})\mathrm{log}(1-\widehat{t}_{i})], 
		\\[1em]
		L_{\mathrm{TIS}} = \frac{1}{M} \sum_{i=1}^{M} |f_{i} - \widehat{f}_{i} |,
	\end{array}
\end{equation}
where $L_{\mathrm{boundary}}$, $L_{\mathrm{topo}}$ and $L_{\mathrm{TIS}}$ represent the loss functions for MLP$_{\mathrm{boundary}}$, MLP$_{\mathrm{topo}}$ and MLP$_{\mathrm{fre}}$, respectively; $M$ represents the minibatch; the upper and lower boundaries of the two BGs are indexed by $j$ $(j = 1,2,3,4)$; $f$ and $\widehat{f}$ denote the real and predicted frequencies, respectively; $t$ and $\widehat{t}$ represent the real and predicted topological properties (using 0 and 1 for the topologically trivial and non-trivial cases, respectively). The optimizer, learning rate, and batch size used for different prediction tasks are summarized in Table \ref{tab:tab1}. Specifically, the optimizer sets the weight decay to 0.01 for predicting the BG frequencies.

\begin{table}[h]
	\centering
	\caption{Hyperparameters of the MLP}\label{tab:tab1}
	\begin{tabular}{cccc}
		\hline
		Prediction target & Optimizer & LR &  Batch size\\
		\hline
		BG boundaries & Adam & 0.05 & 1024 \\
		BG topological properties & Adam & 0.1 & 512 \\
		Frequencies of TISs & Adam & 0.0001 & 32 \\
		\hline
	\end{tabular}
\end{table}

For the purpose of inverse design, we construct a generative network and connect it in series with the trained forward network to form a TNN, as illustrated in Fig. \ref{Fig.VAE_Z_MLP_TNN}(b). The inputs to the generative network are the design targets, including the BG frequencies (linked with MLP$_{\mathrm{boundary}}$), BG topological properties (linked with MLP$_{\mathrm{topo}}$) and TIS frequencies (linked with MLP$_{\mathrm{TIS}}$), while the outputs are the latent vectors. It is important to note that an output latent vector could generate the geometrical configuration of the PnC unit cells with desired wave characteristics. The loss function is defined as follows:
\begin{equation}
	%\label{eq1}
	\begin{split}
		L_{\mathrm{TNN}} = \frac{1}{M} \sum_{i=1}^{M} \sum_{j=1}^{P} |y_{ij} - \widehat{y}_{ij} |, 
	\end{split}
\end{equation}
where $M$ represents the minibatch; $P$ is the number of design targets; $y$ represents the target value, while $\widehat{y}$ denotes the predicted value generated by the forward network. The hyperparameter settings of the generative network for two considered objectives are listed in Table \ref{tab:tab2}. At last, we point out that the DL models are implemented using PyTorch \cite{paszke2019pytorch}, where the training is performed on the workstation with NVIDIA Quadro RTX 4000 (GPU) and Intel$^\circledR$ Xeon$^\circledR$ Gold 6226R (CPU).

\begin{table}[h]
	\centering
	\caption{Hyperparameters of the TNN}\label{tab:tab2}
	\begin{tabular}{cccc}
		\hline
		Design target & Optimizer & LR &  Batch size\\
		\hline
		BG properties & Adam & 0.0001 & 128 \\
		Frequencies of TIS & Adam & 0.001 & 128 \\
		\hline
	\end{tabular}
\end{table}

\section{Results and Discussions} \label{section3}

\subsection{VAE results}

During the VAE training process, the dataset composed of 900,000 original images of PnC unit cells is created and then divided randomly into the training, testing and validation sets in the 8:1:1 ratio. The reconstruction accuracy of the VAE network is calculated by
\begin{equation}
	%\label{eq1}
	\begin{split}
		Acc_{\mathrm{VAE}} = \frac{1}{DN} \sum_{i=1}^{D} \sum_{j=1}^{N} \overline{x_{ij}\widehat{x}_{ij}} \times 100 \%,
	\end{split}
\end{equation}
where $D$ denotes the number of samples; $N$ is the dimension of the design space with $16 \times 16$ pixels; $x$ and $\widehat{x}$ represent the pixel values of the original and reconstructed images (i.e., 0 and 1), respectively. The expression $\overline{x_{ij}\widehat{x}_{ij}}$ indicates whether $x_{ij}$ and $\widehat{x}_{ij}$ are equal, where it takes the value of 1 if they are equal otherwise it is 0. The accuracy rates of the training, validation and testing sets are 98.07$\%$, 97.32$\%$ and 97.33$\%$ respectively. That is, the trained VAE can effectively extract the features of PnC images and thus reconstruct them with high precision. Moreover, the PnC unit cell images with different numbers of scatterers can be effectively reconstructed, meaning that the dataset possesses high degrees of freedom in the geometrical design. 

\begin{figure}[htbp]
	\centering
	\includegraphics[scale=1]{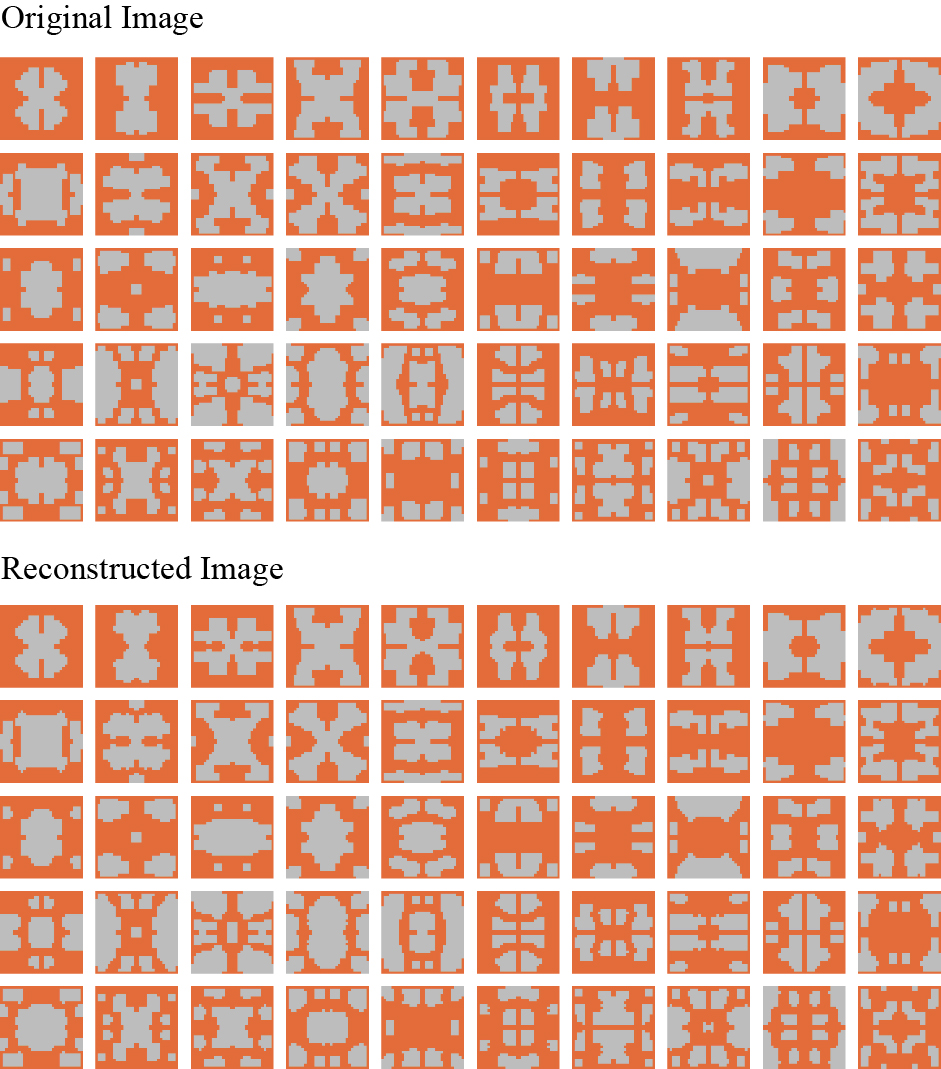}% Here is how to import EPS art
	\caption{\label{Fig.Original_Reconstructed}Original and reconstructed images of PnC unit cells from the validation set for the VAE network, containing different numbers of solid scatterers within a unit cell.}
\end{figure}

According to the number of solid scatterers within in one PnC unit cell, we randomly select 50 examples from the validation set to compare the unit cell images before and after the VAE reconstruction, as displayed in Fig. \ref{Fig.Original_Reconstructed}. It is observed that the images of generated unit cells are almost the same as the original ones, validating the trained VAE for the image reconstruction. As illustrated in Fig. \ref{Fig.Original_Reconstructed}, the images of unit cells contain varying numbers of solid scatterers, signifying the diversity of PnC samples in the dataset. It is noteworthy that, the design of 1D PnCs in this paper could be regarded as the structural topology design, which is different from the shape optimization and the parameter optimization \cite{long2019inverse, zhang2023learning, miao2023deep}.

The histogram in Fig \ref{Fig.VAE_Z_MLP_TNN}(a) depicts the sample distribution in different parameter intervals for each dimension of the latent vector $\textbf{z}$. It is observed that the sample distribution in different parameter intervals is not uniform. If a parameter interval with relatively few samples is used for train the forward network, it is difficult for the forward network to make accurate predictions. Therefore, we restrict the range of the latent vector to $-10 < \mathrm{z}_{i}, (i=1,2,...,10) <10$. Note that the establishment of subsequent datasets are performed within the same parameter range of the latent vector.

\subsection{Prediction of wave characteristics}

\begin{figure}[htbp]
	\centering
	\includegraphics[scale=1]{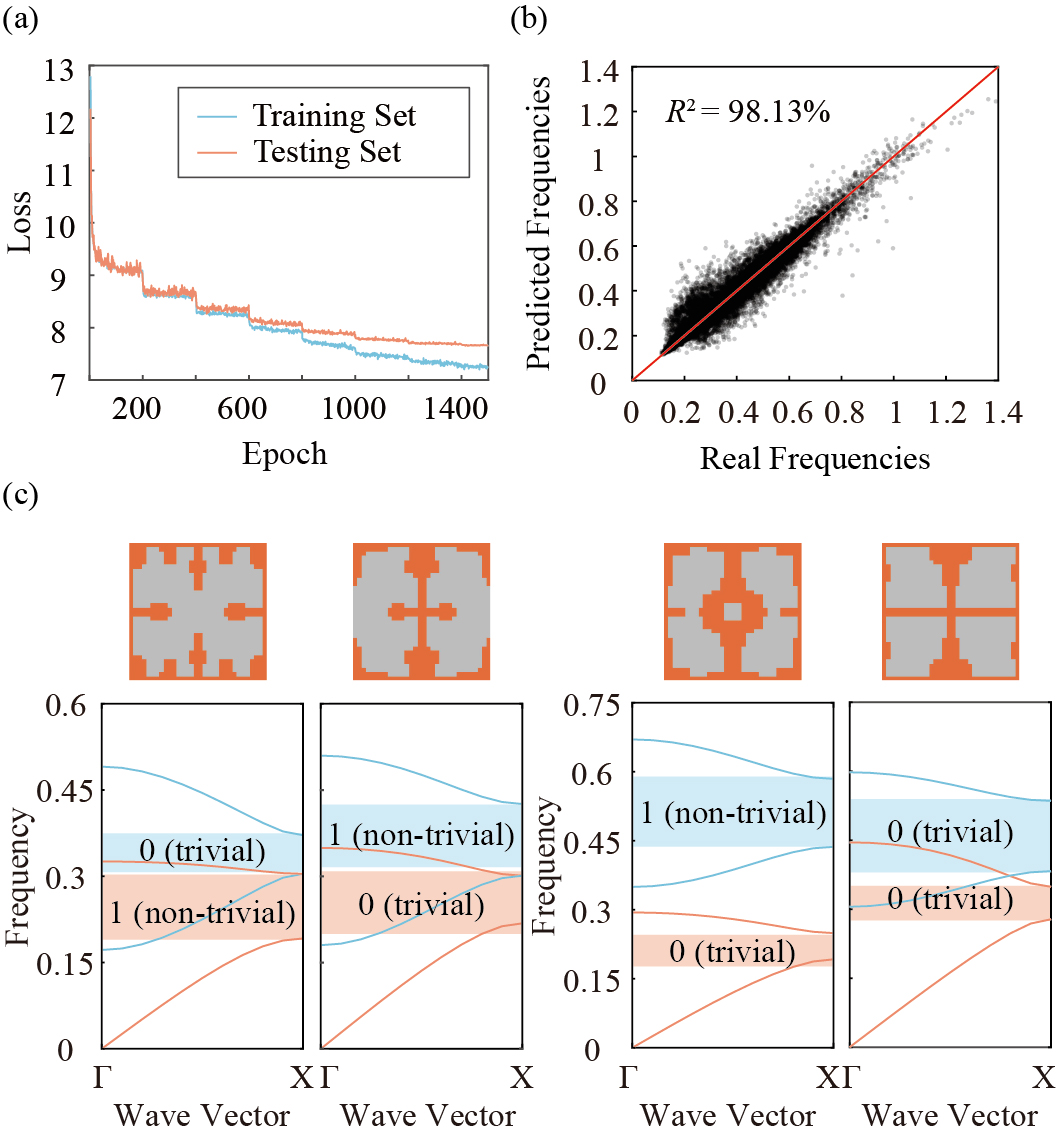}% Here is how to import EPS art
	\caption{\label{Fig.Forward_BG}(a) Training losses of the training and testing sets for predicting the BG frequencies. (b) Scatter plot of the real and predicted frequencies in the testing set. (c) Band structures of four PnC unit cells that are randomly selected from the testing set. The shadows represent the predicted BGs through the MLP and the solid lines depict the results calculated by the FEM, where the orange and blue colors correspond to the symmetric and anti-symmetric modes, respectively.}
\end{figure}

Firstly, a dataset consisting of 20,000 samples of PnC unit cells is generated. Of this dataset, 80$\%$ is randomly allocated to the training set, and 20$\%$ is set aside for the testing set. For predicting the BG frequencies, the training losses of the training and testing sets as a function of epochs are plotted in Fig. \ref{Fig.Forward_BG}(a). Throughout the training process, the LR is reduced by 50$\%$ every 200 epochs. The loss values of both the training and testing sets decrease as the number of epochs increases, stabilizing after 1,200 epochs. The minimal difference between the loss values of the two sets indicates that there is no overfitting during the training process. The effectiveness of the MLP training is characterized by the correlation coefficient between the real and predicted frequencies. The expression of the correlation coefficient is as follows:
\begin{equation}
	%\label{eq1}
	\begin{split}
		R^{2} = \frac{ \sum_{i=1}^{n} (f_{i}-\overline{f}) \times (f_{i}^{'}-\overline{f^{'}})}
		{ \sqrt{ \sum_{i=1}^{n}(f_{i}-\overline{f})^{2}} \times \sqrt{\sum_{i=1}^{n}(f_{i}^{'}-\overline{f^{'}})^{2} }} \times 100 \%,
	\end{split}
\end{equation}
where $\overline{f}$ and $\overline{f^{'}}$ denote the means of the real and predicted frequencies (i.e., $f$ and $f^{'}$), respectively. The correlation coefficients for the training and testing sets are 98.13$\%$ and 98.12$\%$, respectively. The scatterer plot of the real and predicted frequencies in the testing set is shown in Fig. \ref{Fig.Forward_BG}(b), implying that the MLP predicted frequencies agree well with the FEM calculated ones. 

Predicting the topological property of a BG can be treated as a classification problem. Thus, it is necessary to ensure that in the training process the numbers of the PnC samples with trial and non-trivial BGs are similar. However, the established dataset for predicting the BG frequencies does not meet this condition. Additionally, since the topological properties of the BGs for the symmetric and anti-symmetric modes in the same PnC may be different, it is necessary to build the datasets for these two modes separately. To this end, we finally build the datasets for the symmetric and anti-symmetric modes with 68,793 and 80,043 samples, respectively, and the ratio of the samples with distinct topological properties in each dataset is close to 1:1. These datasets are also randomly divided into 80$\%$ for training and 20$\%$ for testing. The network accuracy for predicting the topological property of a BG is assessed using the following function:
\begin{equation}
	%\label{eq1}
	\begin{split}
		Acc_{\mathrm{topo}} = \frac{1}{D} \sum_{i=1}^{D} \overline{ t_{i}\widehat{t}_{i}}  \times 100 \%,
	\end{split}
\end{equation}
where $D$ denotes the number of PnC samples, and $\overline{t_{i}\widehat{t}_{i}}$ indicates whether the real and predicted topological invariants are equal, where it takes the value of 1 if they are equal otherwise it is 0. The accuracy rates of the training and testing sets for predicting the topological invariant of a symmetric mode BG are 99.17$\%$ and 98.75$\%$, respectively. On the other hand, for the anti-symmetric mode, the accuracy rates of the training and testing sets are 97.55$\%$ and 97.12$\%$, respectively. 

Four examples are randomly selected from the testing set to validate the prediction results, as shown in Fig. \ref{Fig.Forward_BG}(c). The shadows represent the predicted BGs and the solid lines depict the band structures calculated by the FEM, where the orange and cyan colors correspond to the results of the symmetric and anti-symmetric modes, respectively. In addition, the topological properties of the BGs are marked in the shaded areas. It is obvious that the predicted results (boundary frequencies and topological properties) are consistent with the FEM results. Moreover, the four PnCs have different numbers of solid scatterers within the unit cells, demonstrating the diversity of PnC samples in the dataset. The average time required to obtain the BG frequencies and the BG topological properties for a single example using the trained forward network is 0.198 $\mu$s and 0.165 $\mu$s, respectively. As a comparison, it takes 24.31 s to calculate the band structure of an PnC example using the FEM. That is, the calculation speed is increased by about seven orders of magnitude. Therefore, for various images of PnC unit cells, we can accurately and rapidly predict the boundaries of the first-order symmetric and anti-symmetric BGs, along with their topological properties.

%It is known that the TIS appears at the interface between two PnCs with distinct topological characteristics: one with a trivial BG and the other with a non-trivial BG.

\begin{figure*}[htbp]
	\Scentering
	\includegraphics[scale=1]{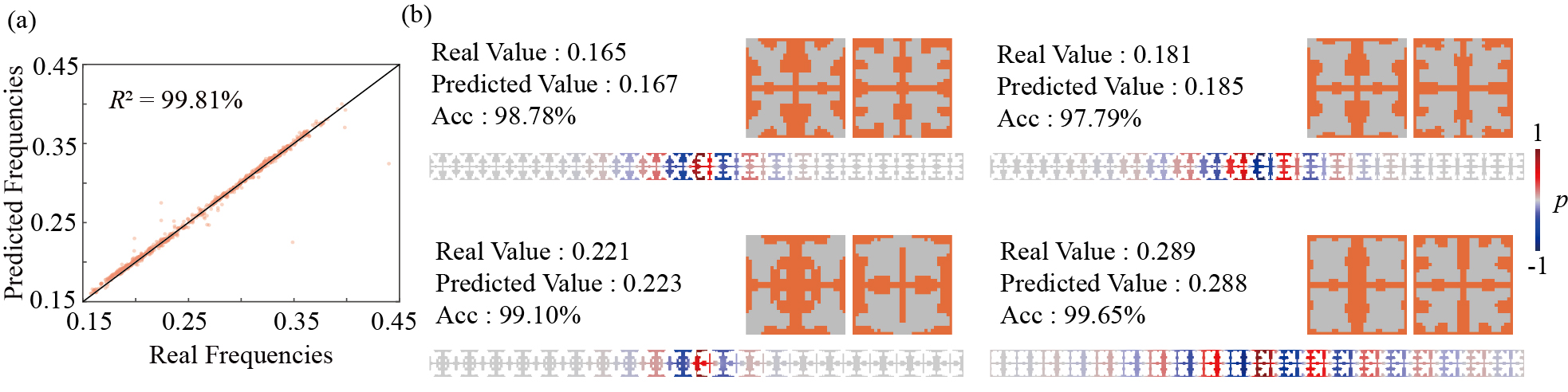}% Here is how to import EPS art
	\caption{\label{Fig.Forward_Topo_S}(a) Scatter plot of 1,000 real and predicted frequencies of the TISs from the testing set for the symmetric mode. (b) Pressure field distributions of four TISs randomly selected from the testing set, along with their real and predicted frequencies.}
\end{figure*}

Also, we explore the potential of the MLP in predicting the TIS frequency, where the latent vectors of two PnC unit cells are employed as the inputs. The datasets for predicting the TIS frequencies within the first symmetric and anti-symmetric BGs are established, where the former and latter contain 33,510 and 28,650 samples, respectively. Here, the symmetric mode is used as an example to present the MLP training results. Figure \ref{Fig.Forward_Topo_S}(a) shows the scatter plot of 1,000 real and predicted TIS frequencies from the testing set. The correlation coefficients for the training and testing sets are 99.81$\%$ and 99.73$\%$, respectively, indicating that the trained MLP network could realize accurate predictions of TIS frequencies. Figure \ref{Fig.Forward_Topo_S}(b) depicts the pressure field distributions of some randomly selected TISs in the testing set. For these interface states, the acoustic pressure field is concentrated at the interface. Also, from the pressure distributions one can notice that these TISs exhibit distinct localization capacity of acoustic waves. This is because different PnC combinations generate distinct BGs. Clearly, the real and predicted TIS frequencies for different PnC combinations exhibit a good agreement, as shown in Fig. \ref{Fig.Forward_Topo_S}(b). The average time required to obtain the TIS frequency for a single example using the trained forward network is 0.191 $\mu$s. As a comparison, the FEM takes 1.74 s to calculate the spectrum of a PnC structure. Moreover, the TIS frequency for the symmetric mode falls within a wide frequency range of $0.15 \sim 0.4$, which can be further customized for different application scenarios. The results of predicting the TIS frequencies for the anti-symmetric mode are provided in \ref{B}.

\subsection{Inverse design}

For either the symmetric or anti-symmetric mode, we propose two goals for the inverse design of 1D PnCs: (i) designing a PnC unit cell with specific BG frequencies and topological characteristics, and (ii) designing a pair of PnC unit cells that supports a specific TIS frequency. This subsection presents the inverse design results of the symmetric acoustic mode, while those of the anti-symmetric mode are provided in \ref{B}.

\begin{figure*}[htbp]
	\Scentering
	\includegraphics[scale=1]{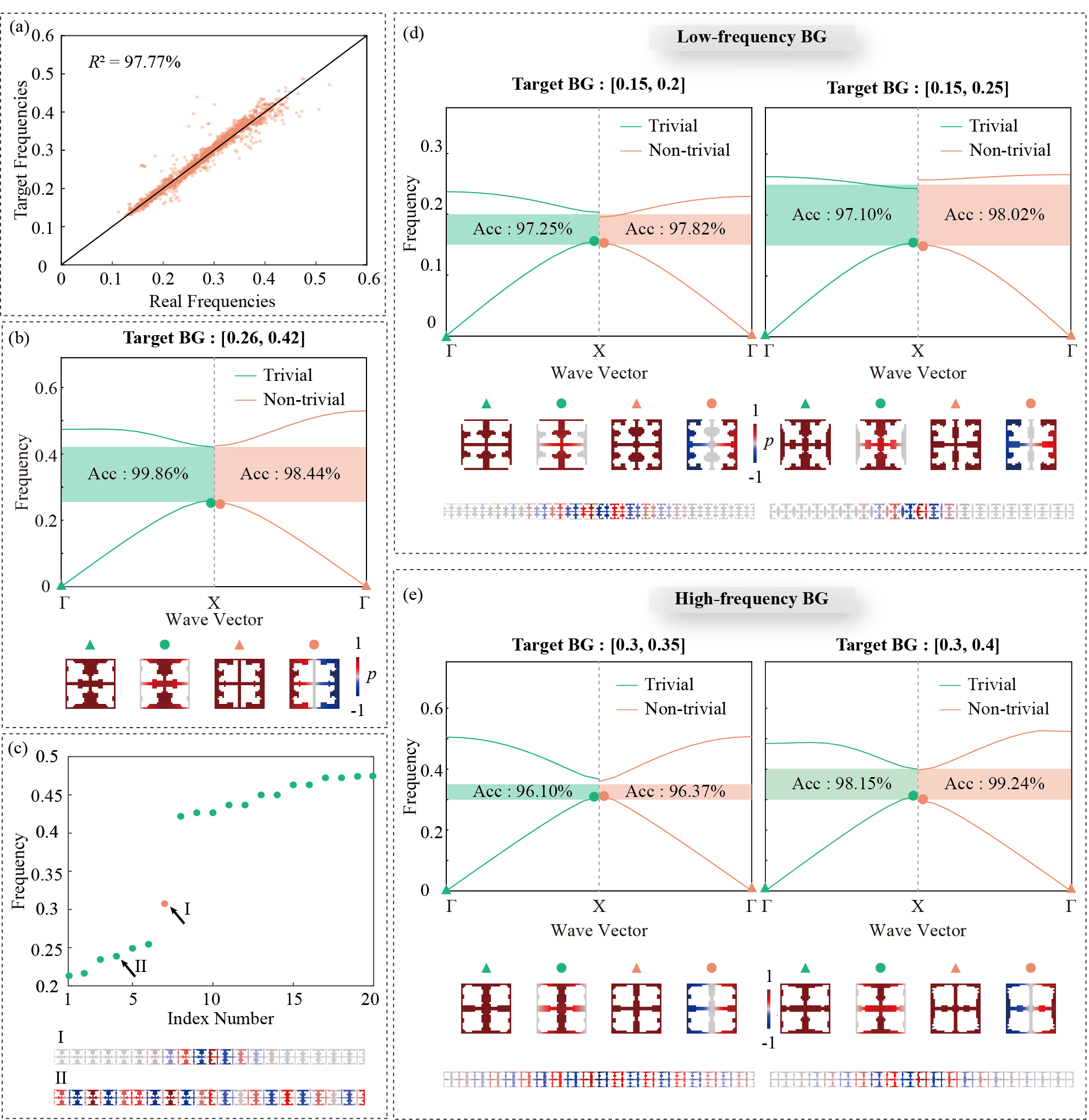}% Here is how to import EPS art
	\caption{\label{Fig.Inver1_S}(a) Scatter plot of 1,000 real and designed BG boundary frequencies from the testing set for the symmetric acoustic mode. (b) An example randomly selected from the testing set: band structures (calculated by the FEM) of the inversely designed PnCs with different topological properties but the same target BG ([0.26, 0.42]). The green and orange colors represent the topological trivial and non-trivial properties, respectively, along with the eigenmodes at points $\Gamma$ and X. (c) Eigenfrequency spectrum of the spliced PnC structure, where the orange and green dots represent the interface and bulk states, respectively. Pressure field distributions of the TIS and bulk state. Inversely designed results targeting the low-frequency (d) and high-frequency (e) BGs, including the band structures of the unit cells and the pressure field distributions of TISs in the spliced PnC structures.}
\end{figure*}

For the first design goal, the frequencies of BG boundaries as well as the BG's topological property serve as the inputs for the generative network. The latent vector is used as the output of the generative network and also the input of the forward network (see Fig. \ref{Fig.VAE_Z_MLP_TNN} for the schematic of the TNN). For a well trained TNN, it is expected that the BG boundaries and the topological property of designed PnC unit cell (represented by the latent vector) match the input targets. From the testing set, 1,000 samples are selected to observe the difference between the real BG properties of designed PnCs (calculated by the FEM) and the design targets. The scatter plot of the real and target BG boundary frequencies are shown in Fig. \ref{Fig.Inver1_S}(a), where $R^{2}$ represents the correlation coefficient of the BG boundary frequencies. On the other hand, the accuracy of the BG topological properties is 1. One example is selected from the testing set to present the results of inverse design. Two PnCs are designed with the same BG frequency range [0.26, 0.42] but with different BG topological properties as the targets, as shown in Fig. \ref{Fig.Inver1_S}(b). The left and right panels are the band structures (calculated by the FEM) of the designed PnCs with the topologically trivial and non-trivial BGs, respectively, and the shadow represent the target BGs. The calculated frequency ranges of the trivial and non-trivial BGs are [0.2601, 0.4210] and [0.2542, 0.4237], respectively, aligning with the design target. In addition, the eigenmodes of the two unit cells at points $\Gamma$ and X are given, confirming the topological properties of the two BGs. By splicing these two PnCs, a PnC structure is created for the generation of a TIS within the BGs. The TIS with the frequency of 0.31 (marked by the orange dot) emerges within the BGs, as plotted in the frequency spectrum [see Fig. \ref{Fig.Inver1_S}(c)]. The acoustic pressure fields for the TIS and one bulk state are also plotted in Fig. \ref{Fig.Inver1_S}(c). For the TIS, the acoustic waves are highly concentrated at the interface between the two designed PnCs with distinct topological characteristics. In contrast, the sound fields of the bulk state are distributed throughout the entire PnC structure.

Furthermore, four sets of target BGs are defined across different frequency ranges. The accuracy of these four sets, the designed structures and the corresponding TISs are displayed in Figs. \ref{Fig.Inver1_S}(d) and \ref{Fig.Inver1_S}(e). During the inverse design process, the average time to design the latent vector for each goal is 0.406 $\mu$s, while decoding the latent vector to generate the unit cell image takes 2.01 ms. In summary, the 1D PnCs that meet the specified objectives can be designed accurately and quickly across a wide frequency range. This also demonstrates the reliability and flexibility of the DL tool for the customized design of PnCs with particular wave characteristics.

\begin{figure*}[htbp]
	\Scentering
	\includegraphics[scale=1]{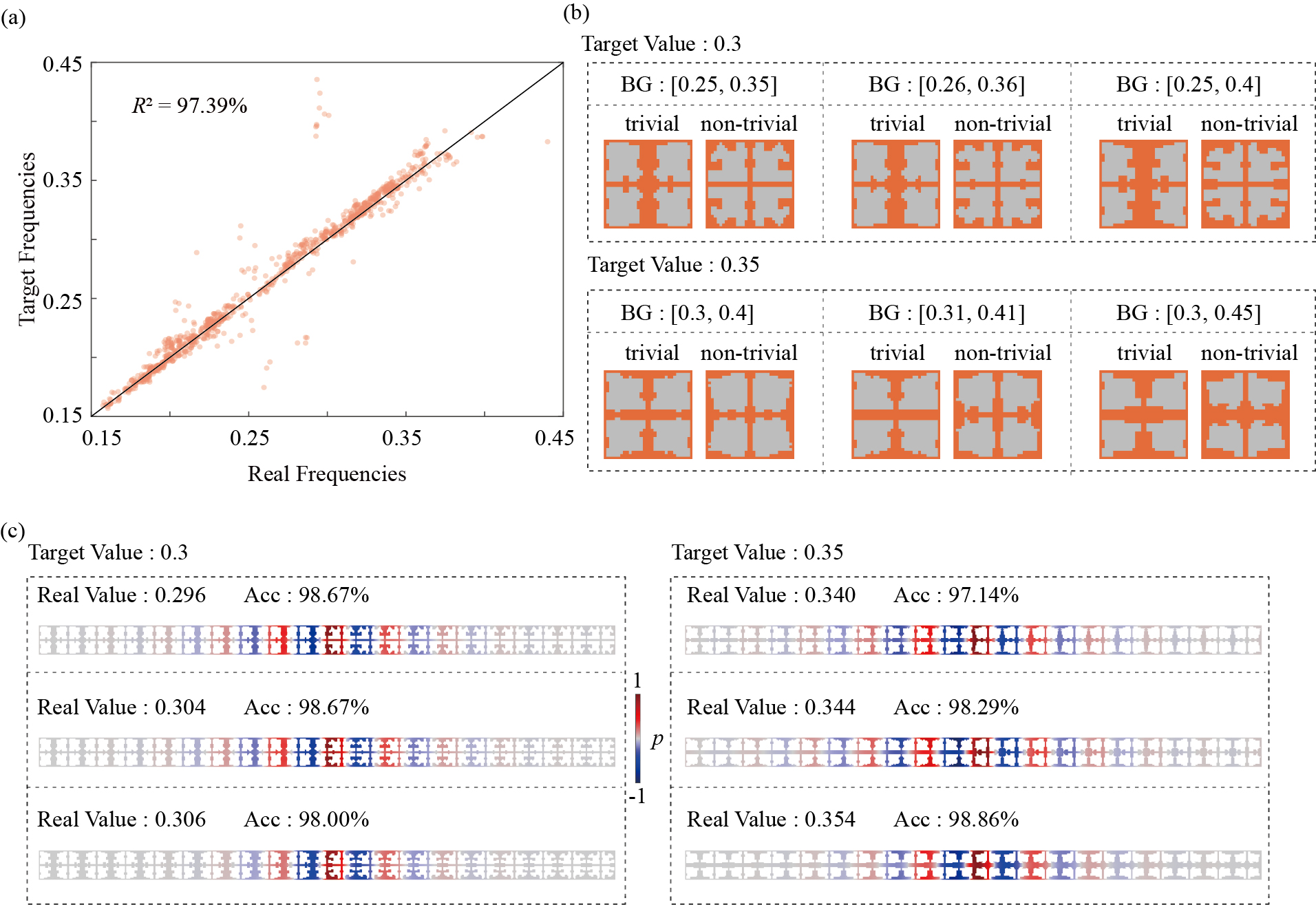}% Here is how to import EPS art
	\caption{\label{Fig.Inver2_S}(a) Scatter plot of 1,000 real and designed TIS frequencies from the testing set for the acoustic symmetric mode. (b) Six sets of inversely designed PnC unit cells. (c) Inversely designed results for two target TIS frequencies, namely 0.3 and 0.35, along with the pressure field distributions of the TISs.}
\end{figure*}

For the second design goal, the presence of similar frequencies corresponding to different PnC structures in the dataset can hinder the network's generalization and lead to the overfitting. To avoid this problem, the TNN inputs both the design target (TIS frequency) and the BG frequency range, as the BGs differ for different PnCs, even if the TIS frequencies are the same. Relatively speaking, we focus exclusively on the TIS frequency. As a result, the weight ratio of TIS frequency to BG range in the loss function is set at 100:1. We select 1,000 samples from the testing set to show the difference between the real TIS frequencies of the designed PnCs (calculated by the FEM) and the target frequencies. The scatter plot of real and target values are presented in Fig. \ref{Fig.Inver2_S}(a). The correlation coefficient for the inverse design results is 97.39$\%$. We provide two target frequencies, 0.3 and 0.35, each associated with three different BG ranges as the inputs. Three sets of unit cell images are designed for each target frequency. Each set consists of both topologically trivial and non-trivial PnCs, as shown in Fig. \ref{Fig.Inver2_S}(b). It is important to note that here we change the inputted BG ranges to obtain different images of PnC unit cells meanwhile the TIS frequency is unchanged. The target and real values, the accuracy of design results, and the pressure fields of the TISs are shown in Fig. \ref{Fig.Inver2_S}(c). Obviously, the sound pressure distributions of the TISs are concentrated at the structural interfaces in the form of a symmetric acoustic mode. It can be seen that the trained TNN accurately performs the inverse design tasks. As the PnC images designed for different BG boundaries vary, a greater disparity results in more pronounced variations in the designed PnC images. In conclusion, a pair of PnCs with any specified TIS frequency can be designed efficiently and accurately via the DL tool, allowing for "one-to-many" designs by adjusting the BG boundary frequencies. 

\subsection{Experimental validation}
\begin{figure*}[htbp]
	\Scentering
	\includegraphics[scale=1]{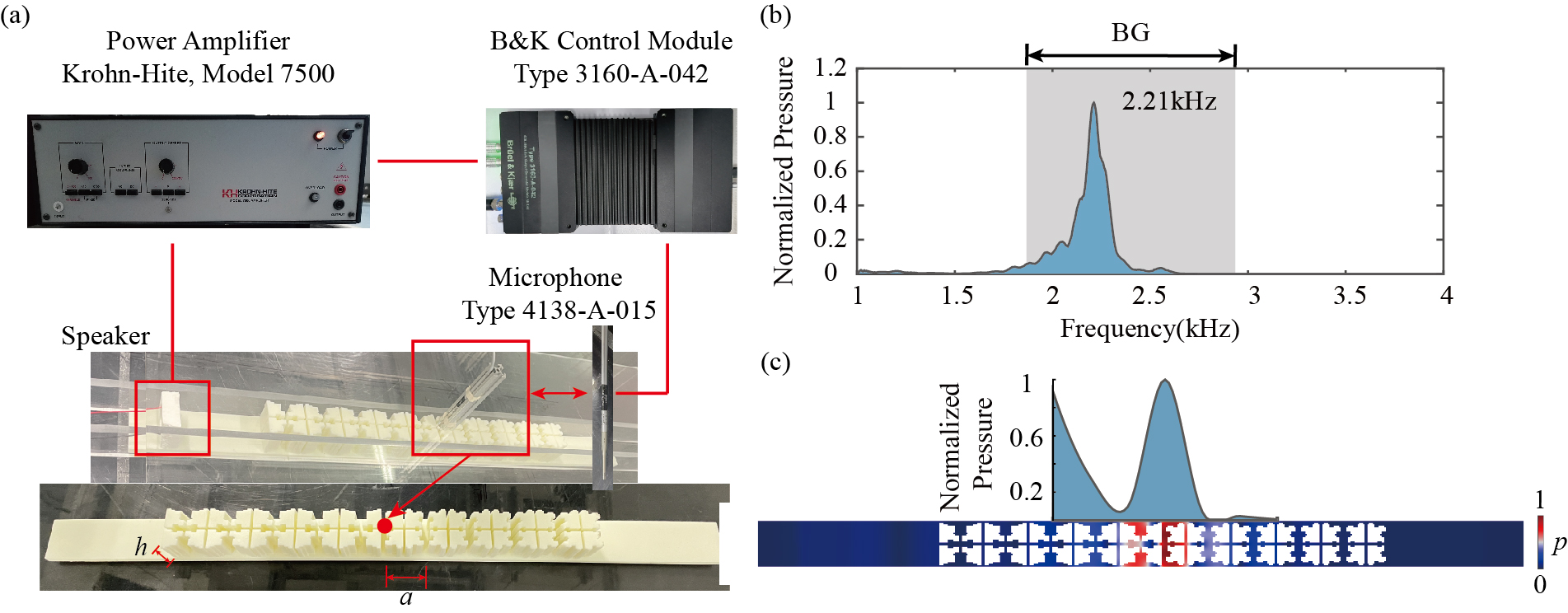}% Here is how to import EPS art
	\caption{\label{Fig.Experiment}(a) Experimental setup for the measurement of acoustic waves and fabricated PnC specimen. (b) Measured acoustic pressure at the structural interface measured within the frequency range of 1 to 4 kHz. (c) Top panel: the measured acoustic pressure at each unit cell center for the excitation frequency of 2.21 kHz. Bottom panel: the numerically calculated distribution of sound pressure field.}
\end{figure*}

For the experimental validation, one designed PnC structure from the testing set is 3D printed, where the lattice constant $a$ is 50 mm and the height $h$ is 40 mm. The experimental setup and the PnC specimen are illustrated in Fig. \ref{Fig.Experiment}(a). The PnC structure consists of a total of 10 unit cells, in which 5 unit cells on the left with the trivial BG and 5 unit cells on the right with the non-trivial one. It is pointed out that PMMA slabs are placed on the lateral sides of the PnC sample as sound hard boundaries, meanwhile on the PnC's top to satisfy the 2D approximation. The loudspeaker is positioned on the left side of the PnC structure as the sound source, while the microphone is placed at the structural interface to measure the acoustic pressure. The sinusoidal signal is generated and post processed using the B$\&$K 3160-A-042 control module.

We utilize the TNN to inversely design a target TIS frequency of 2.13 kHz. The numerical solution obtained by the FEM for the designed PnC structure is 2.2 kHz. The acoustic pressure at the interface is experimentally measured within the frequency range of 1 to 4 kHz, as displayed in Fig. \ref{Fig.Experiment}(b). From the transmission curve one can see that the transmission peak appears at 2.21 kHz, aligning with the numerical simulation. Moreover, we measure the acoustic pressure within the unit cells that are close to the interface at this frequency, as shown in Fig. \ref{Fig.Experiment}(c). The acoustic pressure is strongest at the interface, while pressure on the right side is nearly zero. However, on the left side, the acoustic pressure relatively high due to the setting of the loudspeaker. The distribution of sound pressure fields at this excitation frequency is also numerically calculated, see Fig. \ref{Fig.Experiment}(c). As observed, the experimental results match the numerical calculations, with the acoustic field concentrated at the interface. Therefore, the inverse design capability of the proposed TNN is verified through not only the numerical simulations but also the experiments.

\section{Conclusion} \label{section4}
In this work, the DL approach is employed for the engineering of 1D topological PnCs with a broad design space. We use the VAE to reduce the dimensionality of the topological images of PnC unit cell and extract its key features. The MLP is utilized to accurately predict the first-order BGs of 1D PnCs (i.e., boundary frequencies and topological properties) and the TIS frequencies. The trained MLP is connected to the generative network to form the TNN for the inverse design. The proposed TNN successfully realizes two design goals: (1) the design of PnCs with specific BG properties, including the frequency ranges and topological phases; (2) designing the combination of two PnC unit cells that could generate a TIS with specific frequency. Notably, multiple designs with the same TIS frequency can be obtained by adjusting the inputs. The correlation coefficients for both forward prediction and inverse design exceeds 97$\%$. Moreover, the inverse design of the PnC structure with a specified TIS is experimentally verified, exhibiting a good agreement with the numerical simulation. The DL approach can achieve the design goals accurately and instantaneously. The computational speed is significantly improved by the DL network, where the time for MLP prediction shortens $10^{5} \sim 10^{6}$ times compared to the traditional FEM. More importantly, it only takes several microseconds to perform the inverse design of a PnC geometrical configuration. The TISs in 1D PnCs can be used for energy harvesting, sound filtering, and sensing. Notably, the design domain in this paper has higher degrees of freedom and is more complex, allowing the discovery of new geometries.

\section*{CRediT authorship contribution statement}
Xue-Qian Zhang: Investigation (lead); Methodology (lead); Writing – original draft (lead); Conceptualization (equal). Yi-Da Liu: Investigation (equal); Methodology (equal). Xiao-Shuang Li: Writing – review (equal); Methodology (equal). Tian-Xue Ma: Conceptualization (lead); Supervision (lead); Writing – review (lead); Funding acquisition (equal); Editing (equal). Yue-ShengWang: Funding acquisition (equal); Supervision (equal); Writing – review (equal). Zhou Zhuang: Writing – review (equal).

\section*{Declaration of Competing Interest}
The authors declare that they have no known competing financial interests or personal relationships that could have appeared to influence the work reported in this paper.

\section*{Data availability}
The data that support the findings of this study are available from the corresponding author upon reasonable request.

\section*{Acknowledgments}
This work is supported by National Natural Science Foundation of China (Grant Nos. 12372087, 12021002, 11991031).
%\begin{acknowledgments}
%We wish to acknowledge the support of the author community in using
%REV\TeX{}, offering suggestions and encouragement, testing new versions,
%\end{acknowledgments}
%
\appendix
\section{Calculation of TIS frequencies in finite-sized PnC structures}\label{A}
\begin{figure}[htbp]
	\centering
	\includegraphics[scale=1]{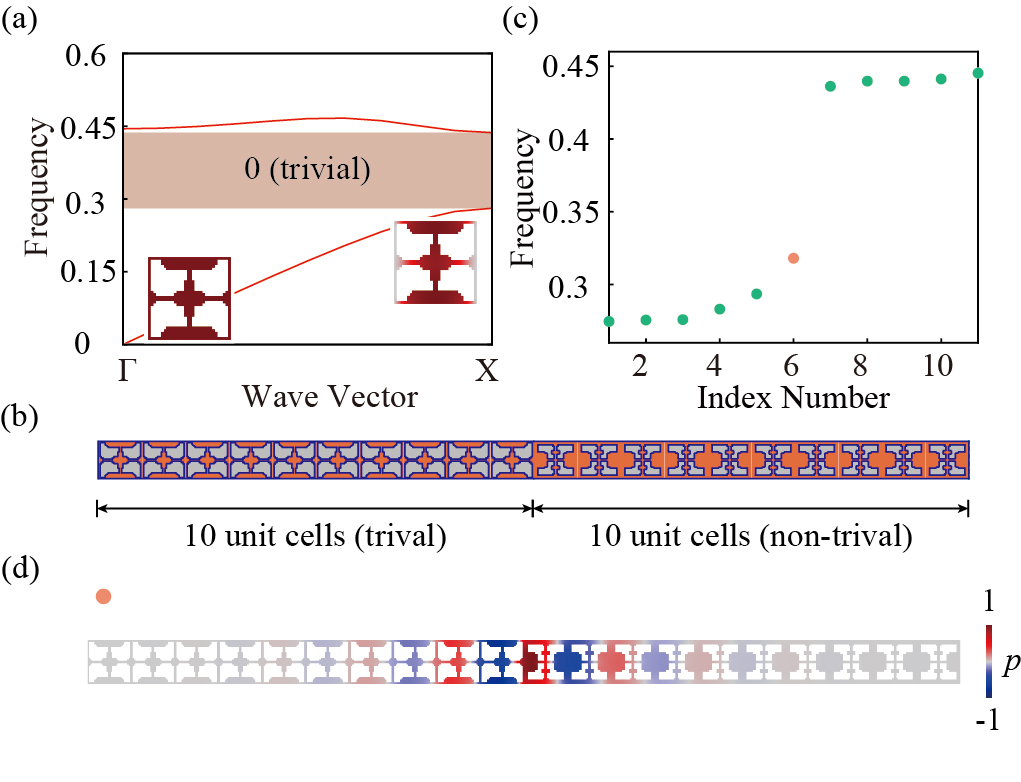}% Here is how to import EPS art
	\caption{\label{Fig.TIS}(a) Band structure of the 1D PnC with a topologically trivial BG, where the BG frequency range is the same as that shown in Fig. \ref{Fig.2}. (b) Finite-sized PnC structure composed of topologically trivial and non-trivial unit cells for the calculation of eigenfrequency spectra. (c) Calculated frequency spectrum of the PnC structure, where the orange and green dots correspond to the interface and bulk states, respectively. (d) Pressure field distribution of the TIS.}
\end{figure}

It is well known that a TIS appears at the interface between two PnCs with distinct topological characteristics: one with a trivial BG and the other with a non-trivial BG. In this work, a finite-sized structure with an interface composed of 20 PnC unit cells is employed to calculate the eigenfrequency spectrum, as illustrated in Fig. \ref{Fig.TIS}(b). Here, we consider an example corresponding to the results in Fig. \ref{Fig.2}. It is proved that the PnC unit cell shown in Fig. \ref{Fig.2} possesses a topologically non-trivial BG. On the other hand, the PnC with the same BG frequency range but different topological characteristics is shown Fig. \ref{Fig.TIS}(a). Note that the topological property can be checked by observing the symmetry of the eigenmodes at highly symmetric points $\Gamma$ and X. The eigenfrequency spectrum of the PnC structure composed of these two kinds of unit cells is displayed in Fig. \ref{Fig.TIS}(c). It can be seen that the TIS appears in the bulk BG. The pressure field distribution of this TIS is provided in Fig. \ref{Fig.TIS}(d), showing a high sound concentration at the interface.

\section{Results of the Anti-symmetric Mode}\label{B}
\begin{figure*}[htbp]
	\Scentering
	\includegraphics[scale=1]{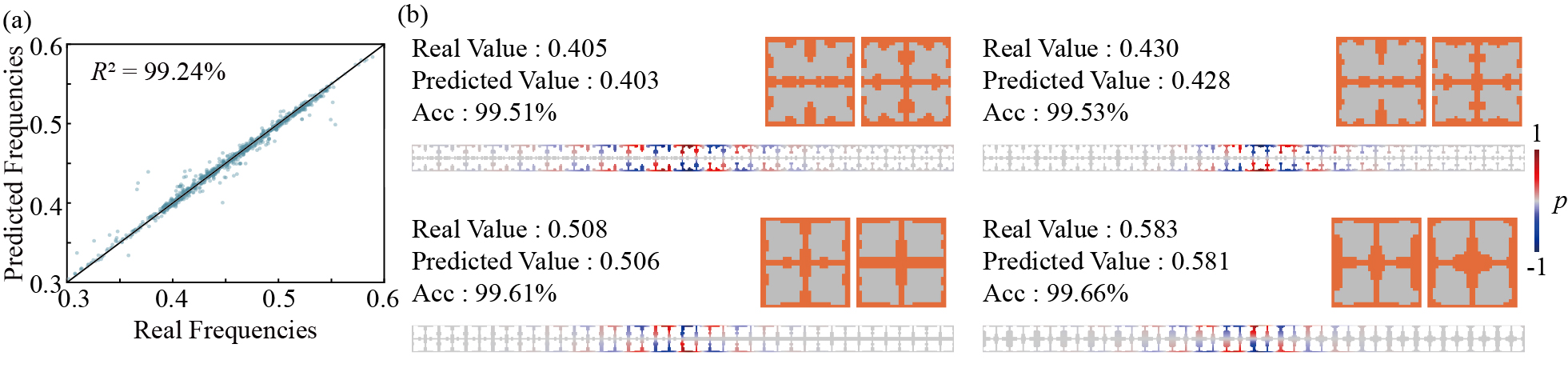}% Here is how to import EPS art
	\caption{\label{Fig.Forward_Topo_A}(a) Scatter plot of 1,000 real and predicted frequencies of the TISs from the testing set for the anti-symmetric mode. (b) Pressure field distributions of four TISs randomly selected in the testing set, along with their real and predicted frequencies.}
\end{figure*}

In this appendix, the results of the acoustic anti-symmetric mode are concerned. Figure \ref{Fig.Forward_Topo_A}(a) shows the scatter plot of 1,000 real and predicted TIS frequencies from the testing set, where the correlation coefficients for the training and testing sets are 99.29$\%$ and 99.24$\%$, respectively. The pressure field distributions of four randomly selected TISs from the testing set are depicted in Fig. \ref{Fig.Forward_Topo_A}(b), along with the real and predicted TIS frequencies for these four combinations of PnC unit cells. The TIS frequency for the anti-symmetric mode falls within a wide frequency range of $0.3 \sim 0.6$. In this frequency range, we are able to accurately predict the TIS frequency of two different PnCs through the trained MLP. In addition, the pressure field of each TIS is concentrated at the junction of the two kinds of crystals, which further verifies the localization characteristics of TISs for the anti-symmetric mode.

\begin{figure*}[htbp]
	\Scentering
	\includegraphics[scale=1]{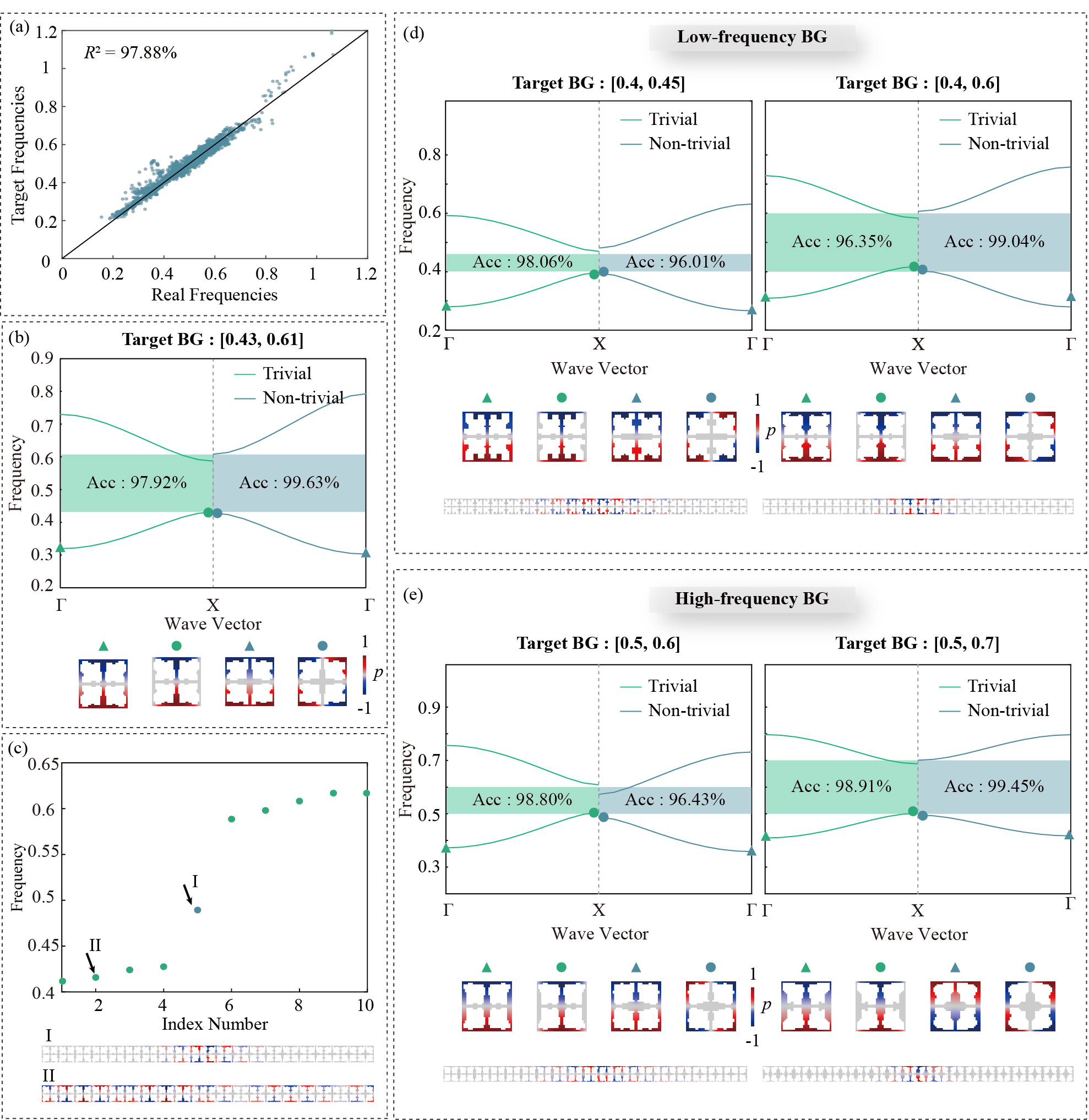}% Here is how to import EPS art
	\caption{\label{Fig.Inver1_A}(a) Scatter plot of 1,000 real and designed BG boundary frequencies from the testing set for the anti-symmetric acoustic mode. (b) An example randomly selected from the testing set: band structures (calculated by FEM) of the inversely designed PnCs with different topological properties but the same target BG ([0.43, 0.61]). The green and blue colors represent topological trivial and non-trivial properties, respectively, along with the eigenmodes at points $\Gamma$ and X. (c) Eigenfrequency spectrum of the spliced PnC structure, where the blue and green dots represent the interface and bulk states, respectively. Pressure field distributions of the TIS and bulk state. Inversely designed results targeting the low-frequency (d) and high-frequency (e) BGs, including the band structures of the unit cells and the pressure field distributions of the TIS in the spliced PnC.}
\end{figure*}

For the first goal of inverse design, we also select 1,000 samples from the testing set to observe the relation between the real and target BG boundary frequencies, see Fig. \ref{Fig.Inver1_A}(a). The correlation coefficient of the BG boundary frequencies and the accuracy of the BG topological properties are 99.24$\%$ and 98.32$\%$, respectively. Additionally, an example from the testing set is designed, in which the BG boundaries of [0.43, 0.61] and different topological properties are considered as the design targets. For the two designed PnCs, the real frequency ranges of the trivial and non-trivial BGs are [0.4316, 0.5869] and [0.4282, 0.6081], respectively, as illustrated in Fig. \ref{Fig.Inver1_A}(b). And the eigenmodes at high symmetry points $\Gamma$ and X, reveal the anti-symmetric modes of the energy bands and the topological properties of the BGs derived from their modal parities. Figure \ref{Fig.Inver1_A}(c) depicts the eigenfrequency spectrum of the spliced PnC structure. It is clearly seen that the TIS emerges within the BG of bulk states. Also, the TIS is demonstrated by its pressure field distribution, which is quite different from that of a bulk state. To further verify the ability of inverse design, we study four design goals across different frequency ranges. It is noteworthy that these designed results are outside the established dataset. The design results for the low-frequency and high-frequency BGs are presented in Figs. \ref{Fig.Inver1_A}(d) and (e), respectively. The real BGs of the designed PnCs vary according to the specified frequency ranges, and all designs are realized by the developed TNN with high accuracy.

\begin{figure*}[htbp]
	\Scentering
	\includegraphics[scale=1]{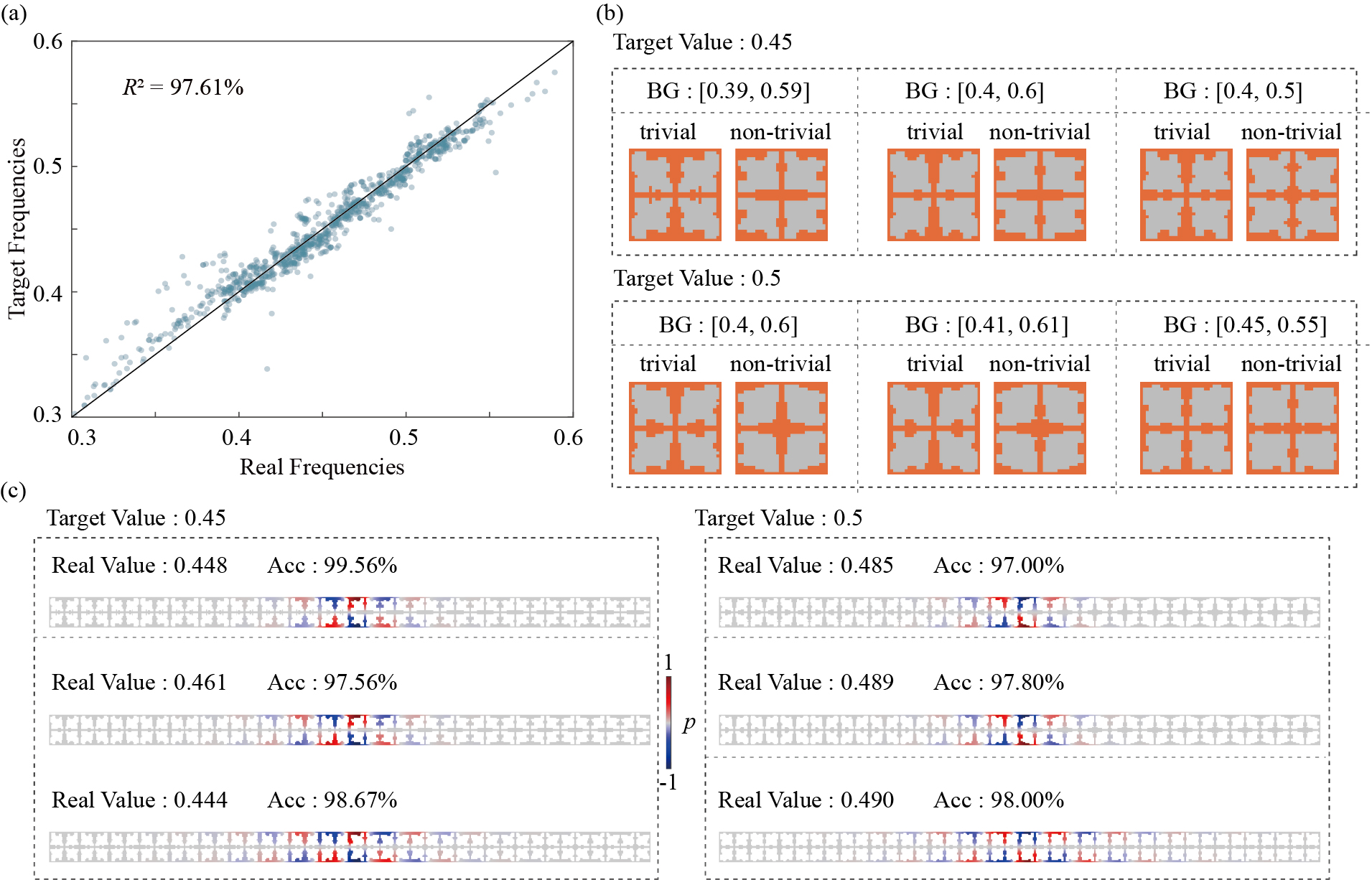}% Here is how to import EPS art
	\caption{\label{Fig.Inver2_A}(a) Scatter plot of 1,000 real and designed TIS frequencies from the testing set for the acoustic anti-symmetric mode. (b) Six sets of iversely designed PnC unit cells. (c) Inversely designed results for two target TIS frequencies, namely 0.45 and 0.5, along with the pressure field distributions of TISs.}
\end{figure*}

For the second design goal, we also provide the scatter plot of real and target TIS frequencies, as shown in Fig. \ref{Fig.Inver2_A}(a). Obviously, the target frequencies are in a good agreement with the real frequencies calculated by the FEM. And two target frequencies, 0.45 and 0.5, each with three different BG ranges are used as the design targets. The results of inverse design are presented in Figs. \ref{Fig.Inver2_A}(b) and (c). For TISs in different frequency ranges, the inverse design can be successfully implemented using the trained TNN. In addition, despite the same TIS frequency, it is possible to design a number of different unit cell images.

\bibliographystyle{elsarticle-num}
\bibliography{ref}% Produces the bibliography via BibTeX.
\end{document}